\documentclass{vienna-conf2019}
\usepackage{graphicx}
\usepackage{hyperref}
\usepackage[]{natbib}  
\usepackage{epstopdf}

\def\BibTeX{{\rm B\kern-.05em{\sc i\kern-.025em b}\kern-.08em
    T\kern-.1667em\lower.7ex\hbox{E}\kern-.125emX}}
\bibpunct{(}{)}{;}{a}{}{,}  

\newcommand{\msun}{\ensuremath{M_{\odot}}}

\newcommand{\rsun}{\ensuremath{R_{\odot}}}

\newcommand{\teff}{\ensuremath{T_{\textrm{eff}}\:}}
\newcommand{\logg}{\mbox{$\log g$}}

\newcommand{\rstar}{\mbox{$R_{\star}$}}

\newcommand{\mstar}{\mbox{$M_{\star}$}}
\newcommand{\massp}{\mbox{$M_{\rm p}$}}
\newcommand{\radp}{\mbox{$R_{\rm p}$}}
\newcommand{\rhop}{\mbox{$\rho_{\rm p}$}}

\newcommand{\re}{\mbox{$R_{\ensuremath{\oplus}}$}}
\newcommand{\me}{\mbox{$M_{\ensuremath{\oplus}}$}}
\newcommand{\fe}{\mbox{$F_{\ensuremath{\oplus}}$}}

\newcommand{\kep}{\mbox{\textit{Kepler}}}

\newcommand{\planet}{HD\,221416\,b}
\newcommand{\tess}{{\it TESS}}
\newcommand{\gaia}{\mbox{\textit{Gaia}}}

\newcommand{\radstar}{\mbox{$2.943\pm0.064$}}
\newcommand{\massstar}{\mbox{$1.212\pm0.074$}}
\newcommand{\agestar}{\mbox{$4.9\pm1.1$}}

\newcommand{\denplanet}{\mbox{$0.431\pm0.062$}}
\newcommand{\incplanet}{\mbox{$343\pm24$}}

\newcommand{\radplanete}{\mbox{$9.17\pm0.33$}}
\newcommand{\massplanete}{\mbox{$60.5\pm5.7$}}

\begin{document}

\TitreGlobal{Stars and their variability observed from space}


\title{Solar-Like Oscillations: Lessons Learned \& First Results from TESS}

\runningtitle{Solar-Like Oscillations}

\author{Daniel Huber}\address{Institute for Astronomy, University of Hawai`i, 2680 Woodlawn Drive, Honolulu, HI 96822, USA}

\author{Konstanze Zwintz}\address{Institut f\"ur Astro- und Teilchenphysik, Universit\"at Innsbruck, Technikerstrasse 25, A-6020 Innsbruck}

\author{the BRITE team}




\setcounter{page}{237}


\maketitle


\begin{abstract}
Solar-like oscillations are excited in cool stars with convective envelopes and provide a powerful tool to constrain fundamental stellar properties and interior physics. We provide a brief history of the detection of solar-like oscillations, focusing in particular on the space-based photometry revolution started by the CoRoT and \kep\ Missions. We then discuss some of the lessons learned from these missions, and highlight the continued importance of smaller space telescopes such as BRITE constellation to characterize very bright stars with independent observational constraints. As an example, we use BRITE observations to measure a tentative surface rotation period of $28.3 \pm 0.5$ days for $\alpha$\,Cen\,A, which has so far been poorly constrained. We also discuss the expected yields of solar-like oscillators from the \tess\ Mission, demonstrating that \tess\ will complement \kep\ by discovering oscillations in a large number of nearby subgiants, and present first detections of oscillations in \tess\ exoplanet host stars.
\end{abstract}

\begin{keywords}
Stars: oscillations, fundamental parameters, Planets and satellites: fundamental parameters
\end{keywords}


\section{Introduction: A Brief History of Solar-like Oscillations}

Solar-like oscillations in cool stars are excited by turbulent convection in the outer layers \citep[e.g.][]{houdek99} and most commonly described by a spherical degree $l$ (the total number of node lines on the surface), azimuthal order $|m|$ (the number of node lines that cross the equator), and radial order $n$ (the number of nodes from the surface to the center of the star). Modes with higher spherical degrees penetrate to shallower depths within the star, and thus the detection of radial ($l=0$) and non-radial ($l>0$) modes provides a diagnostic for the interior structure and fundamental properties of stars. Solar-like oscillators typically exhibit a rich oscillation spectrum with regular spacings, enabling mode identification through simple pattern recognition \citep[see e.g.][for introductory reviews]{bedding11b,aerts20}.

Following the discovery of oscillations in the Sun in the 1960's \citep{leighton62}, early efforts to detect oscillations in other stars focused on ground-based radial-velocity observations. The first confirmed detection of oscillations in a star other than the Sun was made in Procyon by \citet{brown91}, followed by the first 
detection of regularly spaced frequencies in $\eta$\,Boo by \citet{kjeldsen95}. The greatly improved radial velocity precision for detecting exoplanets enabled the detection of oscillations in several nearby main sequence and subgiant stars such as $\beta$\,Hyi \citep{bedding01,carrier01}, $\alpha$\,Cen\,A \citep{bouchy01,butler04} and B \citep{carrier03,kjeldsen05} as well as red giant stars such as $\xi$\,Hya \citep{frandsen02} and $\epsilon$\,Oph \citep{deridder06}. 

\begin{figure}
 \centering
 \includegraphics[width=1\textwidth,clip]{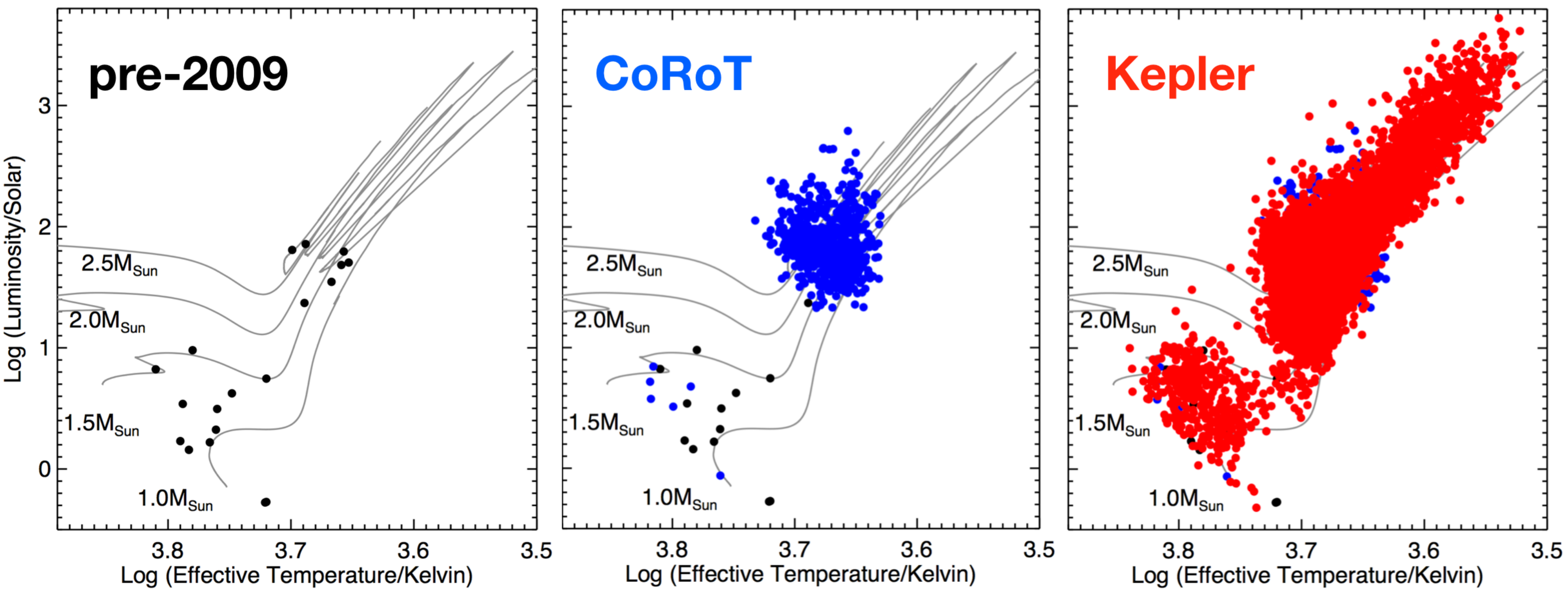}      
  \caption{H-R diagram showing stars with detected solar-like oscillations prior to 2009 (left panel) and after adding detections by the CoRoT (middle panel) and \kep\ (right panel) missions. Grey lines show solar-metallicity evolutionary tracks with masses as marked. The space-photometry revolution has increased the number of solar-like oscillators by three orders of magnitude over the past decade.}
  \label{fig1}
\end{figure}

Some of the first space-based photometric observations of solar-like oscillations were obtained by the Canadian space telescope MOST \citep[Microvariability and Oscillations in Stars,][]{walker03,matthews07}, which initially yielded a non-detection in Procyon \citep{matthews04} but later confirmed a detection that was consistent with radial velocity observations \citep{guenther08,huber11}. MOST also detected oscillations in red giants \citep{barban07}, including observational evidence for non-radial modes \citep{kallinger08b}. Space-based 
observations of solar-like oscillations were also performed using the startracker of the WIRE 
(Wide-Field Infrared Explorer) satellite \citep{schou01,retter03,bruntt05,stello08}, the 
SMEI (Solar Mass Ejection Imager) experiment \citep{tarrant07} and the 
Hubble Space Telescope \citep{edmonds96,gilliland08,stello09b,gilliland11}.
In total, ground and space-based observational efforts prior to 2009 yielded detections in a total of $\sim 20$ stars 
(see left panel of Figure \ref{fig1}). 

A major breakthrough, which is now widely recognized as the beginning of the space photometry revolution of asteroseismology, was achieved by the French-led CoRoT (Convection Rotation and Planetary Transits) satellite. CoRoT detected oscillations in a number of main sequence stars \citep[e.g.][]{appourchaux08, michel08} and several thousands of red giant stars 
\citep[e.g.][]{hekker09} (middle panel of Figure \ref{fig1}). Importantly, CoRoT unambiguously demonstrated for the first time that red giants 
oscillate in non-radial modes \citep{deridder09}, which opened the door for detailed 
studies of the interior structure of red giants \citep[see][for a recent review]{hekker17}.

The \textit{Kepler} space telescope, launched in 2009, completed the revolution of asteroseismology by covering the low-mass H-R diagram with detections. \textit{Kepler} detected oscillations in over 500 main-sequence and subgiant stars \citep{chaplin14b} and over twenty thousand red giants \citep{hekker11c,stello13,yu16}, enabling the study of oscillations across the low-mass H-R diagram (right panel of Figure \ref{fig1}). The larger number of red giants with 
detected oscillations is due to a combination of two effects: First, oscillation amplitudes increase with luminosity \citep{KB95}, making a detection easier at a given apparent magnitude. Second, the majority of  targets were observed with 30-minute sampling, setting an upper limit of $\logg \sim 3.5$ since less evolved stars oscillate above the Nyquist frequency.

\section{Lessons Learned from CoRoT and Kepler}

CoRoT and \textit{Kepler} yielded numerous breakthroughs for solar-like oscillators. One of the most consequential discoveries was that scaling relations for global asteroseismic observables such as the frequency of maximum power, the large frequency separation, and oscillation amplitudes, all of which can be trivially measured from power spectra, are remarkably precise across nearly the entire low-mass H-R diagram \citep[e.g.][]{stello09,huber11, mosser11c}. The use of these scaling relations started the era of ``ensemble asteroseismology'' through the large-scale determination of stellar radii and masses \citep{kallinger09}, paving the way for the now widely successful synergy between asteroseismology and galactic archeology \citep[e.g.]{miglio13b}. Furthermore, the systematic discovery of mixed modes and rotational splittings opened up numerous breakthrough studies of the interior structure and rotation for subgiants and red giants \citep[e.g.][]{beck11,bedding11,mosser14,stello16}.

Space-based observations of solar-like oscillators also uncovered several new challenges. For example, CoRoT and \kep\ showed that mode lifetimes strongly decrease for hot stars, causing an increase in the linewidths which hampers identification of radial and non-radial modes. The ``bloody F star'' problem has been partially addressed through the phase offset $\epsilon$ \citep{white12}, but remains a major obstacle for performing asteroseismology of hot stars. Additionally, the transition of solar-like oscillators to classical pulsators remains poorly understood, and causes major uncertainties when predicting amplitudes and thus detection yields for current and future space-based missions such as \tess\ and PLATO.

\begin{figure}
 \centering
 \includegraphics[width=1\textwidth,clip]{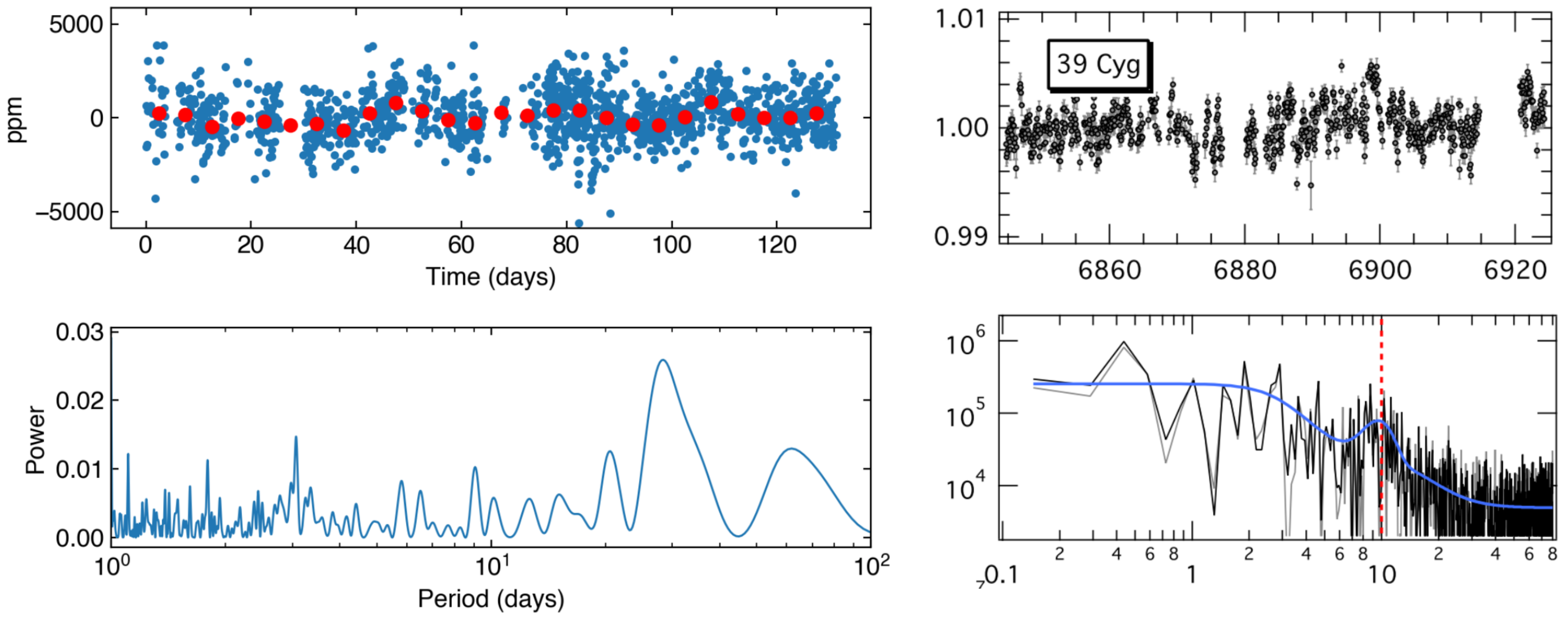}      
  \caption{\textit{Left:} BRITE constellation light curve of $\alpha$\,Cen obtained in 2014 (top), binned into one-orbit (blue circles) and one-day (red circles) averages. A periodogram shows a significant peak at $28.3 \pm 0.5$ days, which may correspond to the rotation period of $\alpha$\,Cen\,A (see text). \textit{Right:} BRITE constellation light curve and power spectrum of the red giant 39\,Cyg, showing the clear detection of solar-like oscillations. From \citet{kallinger19}.}
  \label{fig2}
\end{figure}

Another major challenge for \textit{Kepler} was that the majority of oscillating stars are relatively faint, and thus lack independent observational constraints that are required to fully exploit the information provided by individual frequencies. For example, the potential of the \textit{Kepler} ``legacy'' sample to constrain the convective mixing length parameters \citep{silva17} and initial Helium abundances \citep{verma19} is at times limited by the lack of fundamental constraints such as temperatures, radii and masses from interferometry and/or binary systems.

Small space telescopes such as BRITE constellation play an important role for filling the gap of observing bright very stars. A prominent example is $\alpha$\,Cen: while fundamental properties of both components have been exceptionally well constrained using astrometry and asteroseismology, their rotation periods still remain a matter of debate. Figure \ref{fig2}a shows the BRITE light curve of $\alpha$\,Cen obtained 2014. The continuous coverage over 120 days reveals variability with a period of $28.3 \pm 0.5$ days. $\alpha$\,Cen is not resolved in BRITE observations, but based on the activity cycle of both components \citep{ayres18} the observed period likely corresponds to $\alpha$\,Cen\,A. The period is consistent with but significantly more precise than previous estimates from asteroseismic splittings \citep[$21\pm 9$\,days,][]{fletcher06}, and accounting for the dilution by component B the amplitude of the spot modulation ($\sim$\,370\,ppm) is consistent with relatively quiescent solar-type stars \citep{vansaders19}. BRITE follow-up observations in 2018 provided only a tentative confirmation of this signal, potentially due to change in the spot coverage. Hence the period identified in the 2014 dataset should be viewed with caution.

BRITE has also detected oscillations in bright red giants such as 39\,Cyg \citep[Fig.\ \ref{fig2}, right,][]{kallinger19}. 39\,Cyg ($V=4.4$) is eight magnitudes brighter than the average \kep\ red giant, thus providing an excellent opportunity to study oscillations in red giants with well determined independent parameters.

\section{First Results from the \tess\ Mission}

\subsection{Target Selection}

The NASA \tess\ Mission \citep{ricker14} was launched in April 2018. Located in a 2:1 lunar resonance orbit, \tess\ observes $24 \times 96$ degree fields for 27 days, with continuous coverage near the ecliptic poles. In addition to downloading the entire FOV every 30-minutes (full-frame images, FFIs), \tess\ also observes a subset of targets in 2-minute cadence, which is suitable for the detection of oscillations in solar-type stars.

The selection of asteroseismology targets for the \tess\ prime mission was coordinated within the TESS Asteroseismic Science Consortium (TASC). To select solar-like oscillators, we calculated a detection probability given estimates of effective temperature, luminosity, apparent \tess\ magnitude and the expected number of observed sectors for all stars in Hipparcos and Gaia DR2 following the method by \citet{chaplin11b}, modified for the \tess\ mission. The resulting Asteroseismic Target List (ATL) for the \tess\ mission is described in detail in \citet{schofield19}. 

 \begin{figure}
 \centering
 \includegraphics[width=0.8\textwidth,clip]{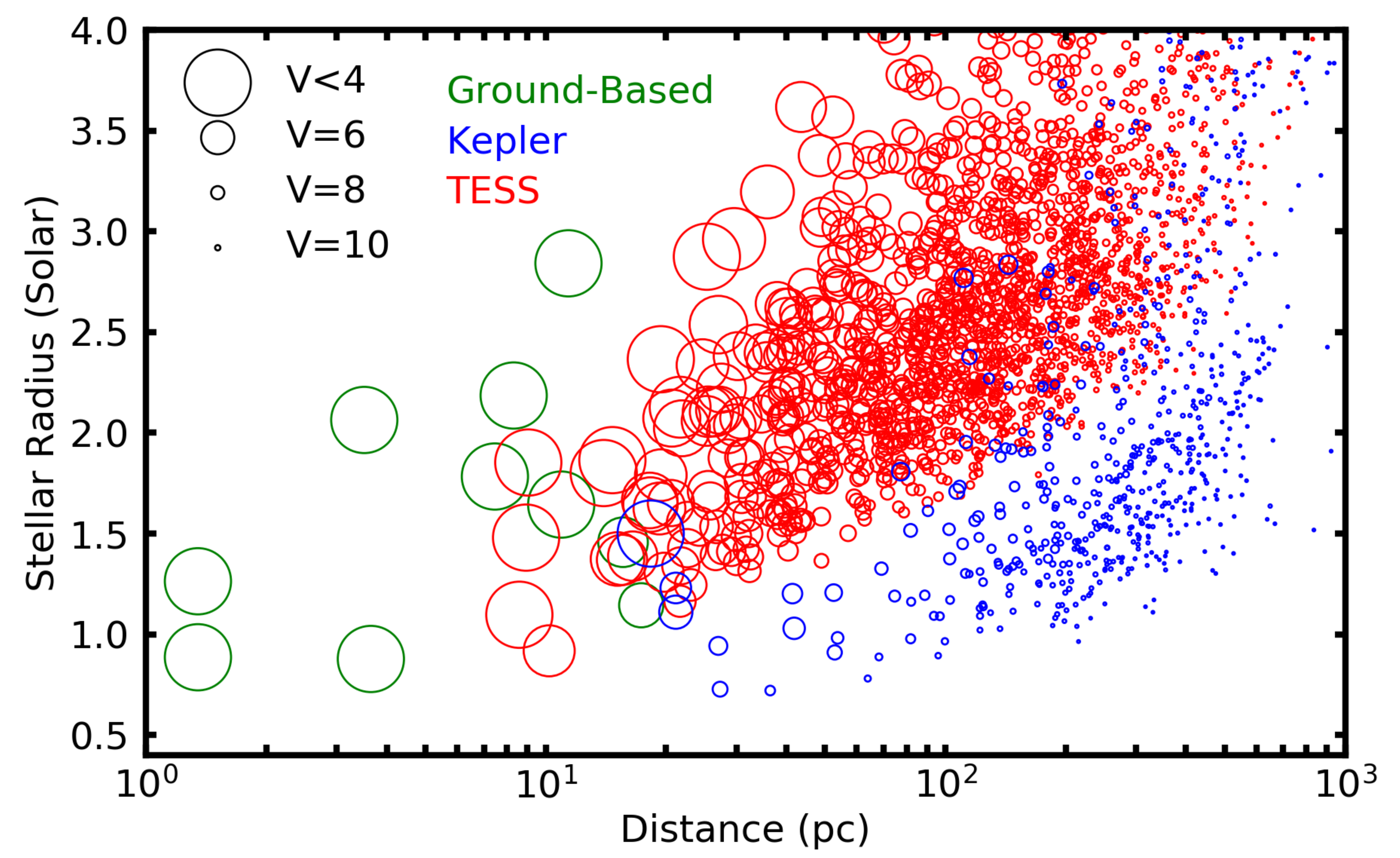}      
  \caption{Stellar radius versus distance for solar-like oscillators detected using ground-based observations (green circles), \kep\ (blue circles), and a representative expected yield from \tess\ (red circles) based on the \tess\ Asteroseismic Target List \citep[ATL,][]{schofield19}. Symbol sizes scale with the apparent V-band magnitude as indicated in the plot. The brightest and closest \kep\ detections are $\theta$\,Cyg \citep{guzik16} and 16\,Cyg\,A and B \citep{metcalfe15}.
\tess\ is expected to complement the \kep\ yield by detecting oscillations in bright, evolved stars.}
  \label{fig3}
\end{figure}

Figure \ref{fig3} shows a representative expected yield of solar-like oscillators from \tess\ compared to ground-based observations and the \kep\ mission. Due to its smaller aperture, the average \tess\ detection is expected to be $\sim$\,5 magnitude brighter, more evolved, and closer compared to \kep. \tess\ is thus expected to complement the parameter space explored by \kep\, which yielded a substantial number of solar-type stars that were relatively faint. Based on preliminary performance the total yield of solar-like oscillators from \tess\ in the prime mission is expected to range between 1000-2000 stars, a 2-4 fold yield increase over the \kep\ mission.

\subsection{Asteroseismology of \tess\ Exoplanet Host Stars}

The search for solar-like oscillations with \tess\ initially focused on exoplanet host stars, for which light curves were made publicly available first to facilitate ground-based follow-up observations. The first claimed detection of oscillations was made for the solar-type star $\pi$\,Men \citep{gandolfi18}, which hosts the first transiting exoplanet discovered by \tess\ \citep{huang18}. Subsequent analysis of the $\pi$\,Men light curve showed that the power spectrum noise level is twice as large as the predicted oscillation amplitude\footnote{\url{https://exofop.ipac.caltech.edu/tess/edit_obsnotes.php?id=261136679}}, thus demonstrating that the claimed detection of oscillations by \citet{gandolfi18} could not have been correct.

\begin{figure}
 \centering
 \includegraphics[width=1\textwidth,clip]{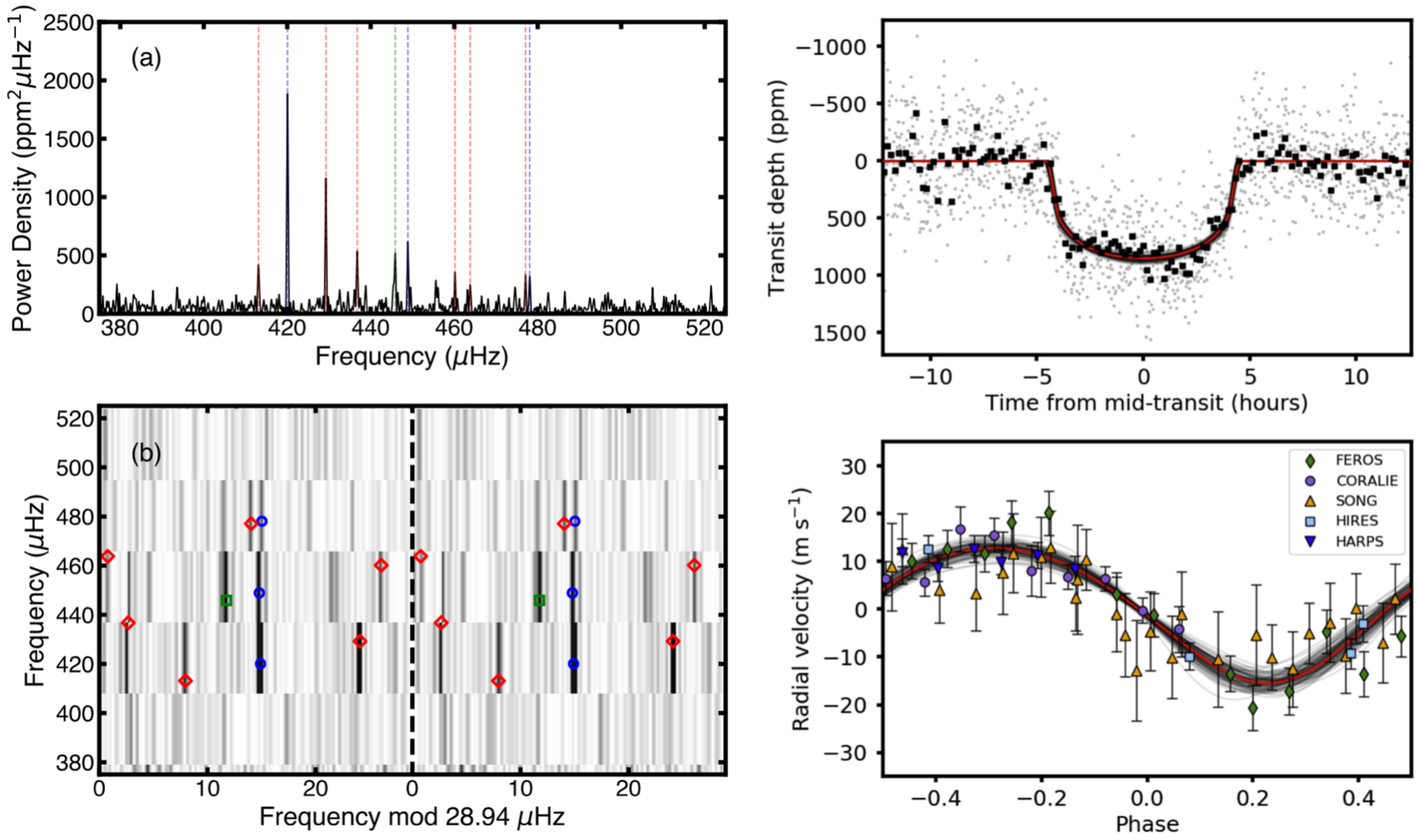}      
  \caption{Detection of solar-like oscillations in HD\,221416 (TESS Object of Interest 197, TOI-197), the first \tess\ asteroseismic exoplanet host star. Left: power spectrum and echelle diagram of the \tess\ time series after removing the planetary transits. Right: Phase folded transit light curve and radial velocity follow-up observations using six different instruments. The combination of asteroseismology, transits and RV measurements constrained the density of the planet to $\sim$\,15\%, making the planet  one of the best characterized Saturn-sized planets to date. From \citet{huber19}.}
  \label{fig4}
\end{figure}

The first confirmed detection of solar-like oscillations by \tess\ was made in the exoplanet host star HD\,221416 (TESS Object of Interest 197, TOI-197), a $V=8.2$\,mag late subgiant star \citep{huber19}. The power spectrum (Figure \ref{fig4}, left) shows a clear detection of mixed dipole modes. Asteroseismic modeling combined with spectroscopic \teff\, metallicity and \gaia\ luminosity yielded a precise characterization of the host star radius ($\rstar = \radstar \rsun$), mass ($\mstar = \massstar  \msun$) and age (\agestar\,Gyr), and demonstrated that it has just started ascending the red-giant branch. The combination of asteroseismology with transit modeling and radial-velocity observations showed that the planet is a ``hot Saturn'' ($\radp= \radplanete \re$) with an orbital period of $\sim$\,14.3 days, irradiance of $F=\incplanet \fe$, moderate mass ($\massp=\massplanete \me$) and density ($\rhop = \denplanet$\,g\,cm$^{-3}$). The properties of \planet\ showed that the host-star metallicity -- planet mass correlation found in sub-Saturns \citep{petigura17} does not extend to larger radii, indicating that planets in the transition between sub-Saturns and Jupiters follow a relatively narrow range of densities. With a density measured to $\sim$\,15\%, \planet\ is one of the best characterized Saturn-sized planets to date.

In addition to stars hosting newly discovered transiting planets, \tess\ has detected oscillations in stars previously known to host planets discovered using the Doppler method \citep[e.g.][]{campante19}.
\tess\ is expected to yield a significant number of new and known exoplanet hosts that are amenable to asteroseismic characterization \citep{campante16}, including new discoveries of transiting planets around oscillating red-giant branch stars \citep[e.g.][]{grunblatt19}.

\section{Conclusions}

Asteroseismology of solar-like oscillators has undergone an exciting revolution over the past decade. In this review I have discussed how small space-based missions such as BRITE Constellation will remain a critical component in characterizing the brightest stars, for example through measuring the poorly constrained rotation period of $\alpha$\,Cen\,A or asteroseismology of bright red giants. Current and future large space-based mission such as \tess\ and PLATO will continue the CoRoT and \kep\ legacy, filling in the parameter space of nearby solar-like oscillators, including the systematic characterization of exoplanet host stars.

\begin{acknowledgements}
It is a pleasure to thank DH's first academic mentor, Werner Weiss, and the whole LOC for a memorable conference in Vienna. We thank Eliana Amazo-Gomez, Tim Bedding, Travis Metcalfe, Bert Pablo, Andrzej Pigulski, and Jen van Saders for discussions on the rotation period of alpha Cen. DH acknowledges support by the National Aeronautics and Space Administration (80NSSC19K0379) through the TESS Guest Investigator Program. Based on data collected by the BRITE Constellation satellite mission, designed, built, launched, operated and supported by the Austrian Research Promotion Agency (FFG), the University of Vienna, the Technical University of Graz, the University of Innsbruck, the Canadian Space Agency (CSA), the University of Toronto Institute for Aerospace Studies (UTIAS), the Foundation for Polish Science \& Technology (FNiTP MNiSW), and National Science Centre (NCN).

\end{acknowledgements}

\bibliographystyle{aa}  
\bibliography{huber_9k04} 

\end{document}